\documentstyle[11pt]{article}
\catcode`\@=11
\begin{document} \openup6pt

\title{Cosmological Models  with Variable Gravitational and Cosmological  constants in $R^{2}$ Gravity}

\author{ P. S. Debnath\thanks{parthasarathi6@hotmail.com } and  B. C. Paul\thanks{bcpaul@iucaa.ernet.in}\\
         Physics Department, P. O. North Bengal University \\
       Siliguri, Dist. : Darjeeling, PIN : 734 430, India}

\date{}

\maketitle

\begin{abstract}

 We consider the evolution of a flat Friedmann-Roberstson-Walker Universe in a higher
derivative theory, including $\alpha \; R^{2}$ 
terms to the Einstein-Hilbert action in the presence of a variable gravitational and cosmological constants. We study here the evolution of the  gravitational and cosmological constants in the presence of radiation and matter domination era of the universe. We present here
 new cosmological solutions which are physically interesting for model building.

\end{abstract}

\vspace{2.5in}
\pagebreak

{\bf I. INTRODUCTION :}

In recent years there has been a growing interest  to study cosmological models with variable cosmological constant [1] and gravitational constant  [2]. The interest stems  from the observational analysis of type Ia supernova with red shift parameter $Z\leq1$ [3]. Supernovae study provides evidence that we may be living in a low mass density  universe of the order of $\Omega \sim 0.3$ [4]. However, inflationary universe model which is an essential ingredient in  modern cosmology predicts, density parameter  $\Omega \sim 1$. Thus a major part of the matter content in the universe remains unobserved, which we call dark matter. It is also understood from the observational analysis of type Ia supernovae that the present universe is accelerating.        This leads to the assumption that there exists a  small positive cosmological constant ($\Lambda$) even at the present epoch. It is also understood in modern Cosmology that our universe might have came through a phase of rapid expansion or inflation in the past. Such a rapid expansion or inflation  may be obtained if one considers an Einstein-Hilbert action with a cosmological constant. To accommodate and to interpret the above facts several physical models are proposed  in recent times with variable cosmological and gravitational constant. The effect of extra dimensions in the evolution of a homogeneous isotropic universe and a first quantitative study on the space and/or time dependence of fine structure and Newton's constants were taken up by Forgacs and Horvath [5].  It was shown that the time and space variation of fundamental constants is a natural consequence of Kaluza-Klein theories. Bertolami [6] studied cosmological model with a time dependent cosmological term. Later, $\ddot{O}$zer and Taha [7]  and others [8] obtained  models of the universe, which are free from  cosmological problems considering a varying cosmological constant in the case of a closed universe. Recently, a number of literature [9-17] appeared which describe various cosmological issues, for example,  the age problem, observational constraints on $\Lambda $, structure formation and gravitational lenses. Ratra and Peebles [10] studied the dark energy content of the universe considering a varying cosmological constant which may be realized using a scalar field. Dolgov [11] analysed the cosmological evolution of a free massless vector or tensor (not gauge) field minimally coupled to gravity and found that cosmological constant exactly cancels out. Sahni and Starobinsky [12] discussed the observational and theoretical aspects of a small positive  cosmological constant term which may be relevant at the present epoch. The origin of such a small $\Lambda$ has been discussed in the framework of field theoretic techniques and by modelling a dynamical $ \Lambda $ term by a scalar field. Recently, Padmanabhan [13]  disscussed several aspects of the cosmological constant both from the cosmological and field theoretic perspectives in order to explain the cosmological constant problem in the framework of the string theory. Vishwakarma [14]  explored different models to estimate the density parameter and/or deceleration parameter for the variable $\Lambda $ models and numerically analysed  the observational data in cosmology.
It is found that the models with dynamical cosmological term ($\Lambda$) are acceptable with observations. Birkel and Sarkar [15] studied the nucleosynthesis and obtained a bound on a time varying $\Lambda$.

In the last few years a number of work reported in cosmology with variable cosmological and gravitational constant with perfect fluid as a source in isotropic [16] as well as in anisotropic models [17] of the universe in the frame work of Einstein gravity. In this paper we study isotropic  cosmological models in a higher  derivative theory (which is also known as a generalized theory)  with variable cosmological and gravitational constant to explore cosmological solutions which can accommodate the recent observations in the universe.
A number of literature appeared in the last few decades on the generalized theory of gravity with additional terms in $R^2$, $R_{\mu\nu\rho\delta}R^{\mu\nu\rho\delta}$, $C_{\mu\nu\rho\delta} C^{\mu\nu\rho\delta}$ in the Lagrangian. However, in 4 dimensions only $R^{2}$ -term is important and the combinations of terms e.g. Gauss-Bonnet combinations occurs as Euler number in the Einstein action. The generalized theory of gravity has a number of good features. It is renormalizable, asymptotically free theory [18]. However, the unitarity  problem of the theory is not yet settled [19].

The paper is organized as follows : in sec. II, we present the gravitational action and field equations, Cosmological solutions are given in sec. III and a brief discussions in sec. IV.

\vspace{.5 in}

{\bf II. GRAVITATIONAL ACTION AND  FIELD EQUATIONS :}

We consider a gravitational action with higher  order terms in the scalar curvature $(R)$ with a variable cosmological constant $ \left( \Lambda(t)  \right) $ and variable gravitational constant $\left( G(t) \right)$ which is given by 

\begin{equation}
{\large I} = - \int  \left[ \frac{f(R)}{16 \pi G(t)}   + L_{m}  \right] \sqrt{- g} \; d^{4}x
\end{equation}
where $f(R)$ is a function of $ R$ and its higher power including cosmological constant  and $ { L_m}$ represents the matter lagrangian.\\ Variation of action (1) with respect to  $g_{\mu\nu}$ yields

\[
f'(R)\; R_{\mu\nu} - \frac{1}{2} \; f(R)\; g_{\mu\nu} + f''(R) \left( \nabla_{\mu} \nabla_{\nu} R -\Box{R} g_{\mu\nu} \right) + 
\]
\begin{equation}
f'''(R)\left( \nabla_{\mu}R \nabla_{\nu} R  - \nabla^{\sigma} R \nabla_{\sigma} R \; g_{\mu\nu}
\right) = - \; 8 \pi G \; T_{\mu\nu}
\end{equation}
where $ \Box = g_{\mu\nu}\nabla^\mu \nabla^\nu $ and $ \nabla_\mu $ is the covariant differential operator, and prime represents the derivative with respect to $ R$, $T_{\mu\nu}$ is the energy momentum tensor for matter determined by ${L_m}$ .

We consider flat Robertson-Walker spacetime given by the metric
\begin{equation}
ds^{2}=-dt^{2}+a^2(t)\left[dr^{2}+r^{2}(d{\theta}^{2}+sin^{2}{\theta } d{\phi}^{2})\right ]
\end{equation}
where $a(t)$ is the scale factor of the universe . The scalar curvature in this case is 
\begin{equation}
 R=- 6  \; [ \dot{H} + 2H^{2}]
\end{equation}
where  $ H=\frac{\dot{a}}{a} $ is the Hubble parameter. The trace and (0,0) components of eq.(2) are given by
\begin{equation}
R f'(R)- 2f(R)+3f''(R) \left(\ddot{R}+3 \frac{\dot{a}}{a} \dot{R} \right) + 3 f'''(R) \dot{R} +8\pi G\ \  T =0,
\end{equation}
\begin{equation}
f'(R)R_{00} +\frac{1}{2} f(R) -3f''(R)\frac{\dot{a}}{a}\dot{R} + 8\pi G \ \ T_{00} =0
\end{equation}
where dot represents derivative with respect to comoving time (t).\\
 Now we consider $ f(R)= R +\alpha R^2 -2 \Lambda(t)$. The equations (5) and (6) yield
\begin{equation}
H^{2}-6{\alpha}\left[2H \ddot{H} -\dot{H}^{2} +6\dot{H} H^{2}\right]
=\frac{8{\pi}G(t) {\rho}}{3}+\frac{\Lambda(t)}{3},
\end{equation}

\begin{equation}
\frac{d\rho}{dt}+3(\rho+p  )H= - \left(\frac{\dot{G}}{G}\; \rho +\frac{\dot{\Lambda}}{8{\pi}G}\right) 
\end{equation}
where $\rho $ is the cosmic fluid energy density and $p $ represents the pressure. The equation of state is
\begin{equation}
 p =(\gamma -1)\rho
\end{equation}
where $\gamma $ is a constant
$(1\leq\gamma\leq2)$. The deceleration parameter $(q)$ is releted to $H$  as 
 $$ q =\frac{d}{dt}\left( \frac{1}{H}\right) - 1$$

{\bf III. COSMOLOGICAL SOLUTIONS :}
\\
We now use  equations $(7)-(9) $  to obtain cosmological solutions. The set of equations contain  six unknown, to solve it we apriori consider time variation of two of them for a given equation of state of the early universe.\\
Let us consider power law behavior of the universe which is given by 
\begin{equation}
 a(t )=a_{0}\; t^{n}
\end{equation}
where $a_{0}$ and $n$ are the constants. We now look for cosmological solutions in two different cases: \ \ (i) Radiation dominated and    (ii) Matter dominated regime of the universe respectively in the next sections .\\ 

{\bf (i) Radiation Dominated Universe :}\\
In this case we have $\gamma =\frac{4}{3}$ i.e. $ p=\frac {1}{3}\rho $ and in the absence of particle creation eq.(8)  can be written as 
\begin{equation}
\frac{d\rho}{dt}+4H \rho =0, 
\end{equation}
\begin{equation}
\dot\Lambda(t)+ 8\pi \rho \dot{G(t)} =0 .\end{equation}
Eq.(11) yields the energy density with the scale factor of the universe, which is
\begin{equation}
 \rho =\frac{\rho_{0}}{a^4}.
\end{equation}
where $\rho_{0}$ is a constant. Using eqs.(7),(12) and (13), one gets evolution of cosmological and gravitatio; S. Perlmutter, {\it Astro. Phys. J.} {\bf 517}, 517 (1999)nal constants, which are  given by
\begin{equation}
 \Lambda(t)=\frac{3n}{2t^2}(2n-1)+ \frac{54\alpha n(n-1)(2n-1)}{t^{4}},
\end{equation}
\begin{equation}
G(t)= \frac{3a_{0}^4}{16\pi\rho_{0}} \left[\frac{n}{t^{2-4n}}+\frac{36\alpha n (2n-1)}{t^{4-4n}}\right].
\end{equation}
The deceleration parameter becomes:\ \ \ \ \ \ $ q = \frac{1-n}{n}$\ \ \ \, which is positive for $n < 1$ and negative for $n > 1$. \\
We note the following cosmological solutions :

(i) When $ n=\frac{1}{2} $ i.e. 
  $  a \sim \sqrt{t}$, we get
$$\Lambda =0 \ \ ; \ \ \ \ \  \ \ \ \ \ \ \ q > 0\ \ $$and
$$G=\frac{3a_{0}^4}{32\pi\rho_{0}}= constant.$$
Thus it is evident that the presence of higher derivative terms do not effect the behaviour of the early universe in radiation dominated universe, in the absence of a cosmological constant and a constant $G$. It has a particle horizon. The      
 cosmological evolution in this case is the same as that obtained for a radiation dominated era in the Einstein gravity. \\
(ii) When $  n=1 $  i.e. $ a \sim t $,  we get
$$\Lambda(t)=\frac{3}{2t^2} \ \ ; \ \ \ \ \ \ \ \ \ \ q = 0 $$and  
$$G(t)=\frac{3a_{0}^4}{16\pi\rho_{0}}\left[t^2+36\alpha \right].$$
In this case $\Lambda(t) $ decreases with time where as $G(t)$ increases. To begin with $G$ is a constant and it is  determined by the coupling constant ($\alpha $), initial size and matter density. For $\alpha < 0$,  the solution is physically relevant for the epoch $t \geq 6 \sqrt{|\alpha |} $. In this case the universe has no particle horizon. \\

We now discuss the solution permitted by the field equations in the presence of  particle creation during radiation domination. The  eq.(8) with radiation becomes 
\begin{equation}
\frac{d\rho}{dt}+4\rho H= - \left(\frac{\dot{G}}{G}\rho +\frac{\dot{\Lambda}}{8{\pi}G}\right). 
\end{equation}
The field eqs.(7) and (16) give the energy density and the cosmological constant which are
 \begin{equation}
\rho =\frac{3n}{16\pi G}\left[\frac{1}{t^2}+\frac{36\alpha (2n-1)}{t^4}\right]
\end{equation}
and
\begin{equation}
\Lambda =\frac{3n}{2 t^2}(2n-1)+\frac{54\alpha n (2n-1)(n-1)}{t^4}
\end{equation}
We now discuss special cases :\\
(i) When $ n=\frac{1}{2}$, we get
$$ \Lambda =0  ,$$
$$\rho = \frac{3}{32\pi G}\frac{1}{t^2}$$
Here the solution  is similar to the usual matter dominated universe in the Einstein gravity, which we get even if $
\alpha \neq 0$ in the absence of cosmological constant.\\
(ii) When $ n=1$, we get
$$\Lambda =\frac{3}{2t^2}  ,$$
$$\rho =\frac{3}{16\pi G} \left[\frac{1}{t^2}+\frac{36\alpha}{t^4}\right]$$
For  $n=1$, the cosmological constant is independent of the higher derivative term, however,  the  energy density is determined by a coupling parameter $\alpha $ of the higher derivative theory. For $\alpha < 0$, physically relevant solutions are obtained for $t\geq 6\sqrt {|\alpha|}$. The above solutions are permitted with a gravitational constant which remains invariant. \\

{\bf (ii) Matter Dominated Universe:}\\
In this case we consider  $ p=0 $ i.e. $\gamma =1 $, and in the absence of particle creation eq.(8) can be written as two decoupled equations, given by
\begin{equation}
\frac{d\rho}{dt} + 3H\rho = 0,
\end{equation}
\begin{equation}
\dot\Lambda(t) +8\pi\rho\dot{G(t) }=0 .
\end{equation}
Eq.(19) yields the energy density
\begin{equation}
\rho =\frac{\rho_{0}}{a^3}
\end{equation}
where $ \rho_{0} $ is a constant. Using eq.(7),(20) and (21), the cosmological constant and gravitational constant are obtained, which are
\begin{equation}
\Lambda(t) =\frac{n}{t^2}(3n-2)+\frac{18\alpha n (2n-1)(3n-4)}{t^4} ,
\end{equation}
and
\begin{equation}
G(t) =\frac{a_{0}^3}{4\pi \rho_{0}} \left[ \frac{n}{t^{2-3n}} + \frac{36\alpha n (2n-1)}{t^{4-3n}} \right] .
\end{equation}
We note the following cosmological solutions :\\
(i) When $  n=\frac{1}{2} $ i.e. $ a \sim \sqrt{t} $, we get
$$ \Lambda(t) =-\frac{1}{4t^2} ; \ \ \ \ \ \ \ \ \ \ q > 0 $$ and
$$ G(t) =\frac{a_{0}^3}{8\pi \rho_{0} \sqrt{t}} $$
In this case one gets solution with negative cosmological constant , however, the Gravitational constant remains positive and decreasing with time. Thus a comparatively less expansion in the matter dominated regime is obtained in this case ( since $a(t) \sim t^{\frac{2}{3}}$, in Einstein gravity).\\
(ii) When $ n=\frac{2}{3}$ i.e.  $ a \sim t^{\frac{2}{3}}$,  we get
$$ \Lambda(t) =-\frac{8\alpha}{t^4} ; \ \ \ \ \ \ \ \ q > 0 $$and
$$G(t) =\frac{a_{0}^3}{4\pi \rho_{0}}\left[ \frac{2}{3}+\frac{8\alpha}{t^2} \right]$$
We obtain Matter dominated solution with $ \Lambda < {0} $ if $ \alpha >{0} $ and $ \Lambda > {0}$ if $\alpha < {0}$, however G attains a constant value at a later epoch, i.e. at $ t\rightarrow \infty$. It has a particle horizon.  This solution is new and interesting in model building for describing late universe. The cosmological constant becomes small depending upon time and the coupling parameter at late epoch. Here one recovers the matter dominated solution of Einstein gravity for $\alpha =0$,( as $\Lambda =0 $ and $ G = G_0$). \\
(iii) when $  n=\frac{4}{3} $, i.e. $ a \sim t^{\frac{4}{3}}$, which is a rapid power law expansion. In this epoch, we get 
$$\Lambda(t) = \frac{8}{3t^2} ; \ \ \ \ \ \ \ \ \ \ q < 0 $$ and 
$$  G(t)  = \frac{a_{0}^3}{4\pi \rho_{0}}\left[80\alpha +\frac{4}{3}t^2\right]$$
Thus in the matter dominated case an accelerating universe results in the presence of a decaying cosmological constant, but with an increasing gravitational constant $G(t)$ for $\alpha > 0$. For $\alpha <0 $, at $ t\rightarrow  0$, one gets solution with  $G(t)$  negative which later attains positive values for  $t > \sqrt {|60\alpha|}$ and thereafter $G(t)$ grows. The solution has no particle horizon. 

In the presence of particle creation and matter domination era, eq.(8) becomes 
\begin{equation}
\frac{d\rho}{dt}+3H\rho= - \left(\frac{\dot{G}}{G} \; \rho +\frac{\dot{\Lambda}}{8{\pi}G}\right). 
\end{equation}
From eqs.(7) and (24), the energy density and the cosmological constant are obtained, which are
\begin{equation}
\rho =\frac{n}{4\pi G}\left[\frac{1}{t^2}+\frac{36\alpha (2n-1)}{t^4}\right] ,
\end{equation}
\begin{equation}
\Lambda =\frac{n}{t^2}(3n-2)+\frac{18\alpha n (2n-1)(3n-4)}{t^4}  .
\end{equation}
 We now discuss special cases :\\
(i) when $  n=\frac{1}{2} $ i.e. $ a \sim \sqrt{t} $, the cosmological parameter and the energy density varies as,
$$ \Lambda =-\frac{1}{4t^2}, $$
$$\rho = \frac{1}{8\pi G t^2}. $$
(ii) When $ n=\frac{2}{3}$ i.e.  $ a \sim t^{\frac{2}{3}}$,  we get
$$ \Lambda =-\frac{8\alpha}{t^4},$$
$$\rho =\frac{1}{6\pi G} \left[\frac{1}{t^2}+\frac{12\alpha}{t^4}\right].$$ 
Thus the usual matter dominated solution is recovered with $\Lambda < 0$ here.\\
(iii) when $  n=\frac{4}{3} $, i.e. $ a \sim t^{\frac{4}{3}}$, which is a rapid power law expansion. In this epoch, the cosmological constant and the energy density become
$$\Lambda = \frac{8}{3t^2}  ,$$
$$\rho =\frac{1}{3\pi G} \left[\frac{1}{t^2}+\frac{60\alpha}{t^4}\right]$$For $n=\frac{1}{2}$ and $n=\frac{4}{3}$; the cosmological constant is independent of the higher derivative term, however, in the later  case  energy density also depends on the coupling parameter associated with the higher derivative term. In the later case it has no particle horizon. But in  the case of $n=\frac{2}{3}$, both the cosmological constant and  energy density is determined by $\alpha$. It may be pointed out here that the gravitational constant $G$  becomes a constant in this case. The solution obtained here describes a universe which is still accelerating in the matter dominated phase. \\
{\bf{IV. Discussions : }}\\
In this paper we investigate isotropic cosmologies with varying $G $ and $\Lambda$ in a $R^{2}$ theory. We studied both radiation dominated and matter dominated phases of evolution of the universe  with varying $G$  and $\Lambda$ and obtained some new cosmological solutions, which may be important to describe a late universe. We also discussed relevance of some of the solutions as  special cases. 
We obtain here some solutions  where $G$ is increasing  similar to that  found by Abdel Rahaman [2] and Arbab  [20] in Einstein gravity. The cosmological constant $\Lambda$,  however, decreases with time in those cases. In the frame work of $R^2$ theory we note an interesting solution during  matter dominated regime. The gravitational constant $G$ decreases initially and   ultimately it attains a constant value at a later stage of evolution of the universe. The  universe evolves as $a \sim t^{\frac{2}{3}}$   which is similar to that one gets in the Einstein gravity during matter domination but we note a realistic variation for the parameter $G$. In this scenario $G $ may be large in the early universe, however,  at the present epoch it becomes a constant. The above solutions are permitted either with (i)
{$\Lambda > 0 $ if $\alpha < 0 $} or (ii) with {$\Lambda < 0 $ if $ \alpha > 0 $} . We also note that in the matter dominated era an interesting new solution exists in $R^2$ theory which admits an accelerating universe (supported by supernova observation), with a decaying cosmological constant and  a constant $G$. It may be pointed out here that an  inflationary solution 
 $(a(t) \sim e^{H_{0} t}) $ are permitted in $R^{2}$ -theory if one considers a non zero $\Lambda$ even with $\rho = p = 0$.  Power law inflation is also admitted in the matter dominated universe (MDU) or in the radiation dominated universe (RDU) with varying $\Lambda $ and $G $.  In MDU, power law expansion  of the universe is permissible if one allows $G$ to be increased at a later stage. However, when $\Lambda \rightarrow  0$, the solution is no longer valid and the usual matter dominated universe solution may be recovered with a constant $G$. The stability analysis of the cosmological solutions discussed here will be taken up elsewhere. \\

\vspace{0.5 in}

$ {\it Ackknowledgements :} $\\
The authors would like to thank $ IUCAA \; Reference \; Centre $ at North  
 Bengal University for providing facilities during the work.
BCP would like to thank University Grants Commission, New Delhi for awarding Minor Research Project (Project No. 30-16/2004(SR)) for completing the work. 
\pagebreak

\end{document}